\documentclass[11pt,a4paper]{article}
\pdfoutput=1

\usepackage[
top    = 2.50cm,
bottom = 2.50cm,
left   = 2.50cm,
right  = 2.50cm]{geometry}
\usepackage{jheppub}
\usepackage[sort&compress]{natbib}

\usepackage{amssymb}
\usepackage{graphicx}
\usepackage{slashed}
\usepackage{amsmath}
\usepackage{hyperref}
\usepackage{verbatim}

\usepackage{caption}
\usepackage{subcaption}

\newcommand{\SL}{\mathrm{SL}}
\newcommand{\SO}{\mathrm{SO}}

\title{Comments on the Random Thirring Model}

\author[a]{Micha Berkooz}
\author[b]{\!, Prithvi Narayan}
\author[c]{\!, Moshe Rozali}
\author[d,e]{\!, Joan Sim\'on}
\affiliation[\,a]{
\it{Department of Particle Physics and Astrophysics,\\
Weizmann Institute of Science, Rehovot 7610001, Israel}}
\affiliation[\,b]{
\it{International Centre for Theoretical Sciences, Hesaraghatta,\\ 
Bengaluru North, 560 089, India}}

\affiliation[\,c]{
\it{
Department of Physics and Astronomy, University of British Columbia,\\
Vancouver, BC V6T 1Z1, Canada}}

\affiliation[\,d]{
\it{School of Mathematics and Maxwell Institute for Mathematical Sciences,\\
        University of Edinburgh, King's Buildings,
        Edinburgh EH9 3FD, UK}}
    
  \affiliation[\,e]{
  	\it{Perimeter Institute for Theoretical Physics, Waterloo, ON N2L 2Y5, Canada}}  
\emailAdd{Micha.Berkooz@weizmann.ac.il}
\emailAdd{prithvi.narayan@gmail.com}
\emailAdd{rozali@phas.ubc.ca}
\emailAdd{j.simon@ed.ac.uk}

\abstract{The Thirring model with random couplings is a translationally invariant generalisation of the SYK model to 1+1 dimensions, which is tractable in the large N limit. 
We compute its two point function, at large distances, for any strength of the random coupling. For a given realisation, the couplings contain both irrelevant and relevant marginal operators, but statistically, in the large N limit, the random couplings are overall always marginally irrelevant, in sharp distinction to the usual Thirring model. We show the leading term to the $\beta$ function in conformal perturbation theory, which is quadratic in the couplings, vanishes, while its usually subleading cubic term matches our RG flow.}

\begin{document}

\maketitle

\section{Introduction}

The Sachdev-Ye-Kitaev (SYK) model \cite{Sachdev:1992fk,KitaevTalks,Polchinski:2016xgd,Maldacena:2016hyu},
\begin{equation}
  S_{\text{SYK}} = \int dt \left(\sum_j \frac{i}{2}\psi_j\partial_t \psi_j - i^{q/2}\sum_{i_1 \dots i_q} J_{i_1\dots i_q} \psi_{i_1}\dots \psi_{i_q} \right)\,,
\end{equation}
 is a quantum mechanical model describing $N$ Majorana fermions $\psi_i$ interacting via a non-linear potential involving gaussian random couplings $J_{i_1\dots i_q}$. It is solvable in the large $N$ limit, exhibits approximate conformal symmetry in the IR, and saturates the quantum chaos bound \cite{Shenker:2013pqa, Shenker:2013yza,Shenker:2014cwa,Maldacena:2015waa,Reynolds:2016pmi}. It has been proposed as the holographic dual of 2d dilaton gravity based on the observation that both sides of the correspondence share the same Goldstone mode effective action, written in terms of the Schwarzian derivative \cite{Almheiri:2014cka,Sachdev:2015efa,Maldacena:2016upp,Jensen:2016pah,Sekino:2008he,Engelsoy:2016xyb,Cvetic:2016eiv}.

In this paper, we consider a 1+1 dimensional generalisation of the SYK model, by considering the Thirring model with random couplings. i.e.
\begin{equation}
\label{IntroMod}
S ={1 \over 2  {\bf \pi} }  \int d^2z \ \left[ \sum_i \left({\bar\nu}^i \partial_z {\bar\nu}^i  + \nu^i  \partial_{\bar z}  \nu^i\right)   + \sum_{i<j,k<l}J_{ij;kl}\nu^i\nu^j{\bar\nu}^k{\bar\nu}^l\right]\,.
\end{equation}
with random couplings $J_{ij;kl}$ drawn from a gaussian ensemble with standard deviation $J$ (we also discuss generalisations of this model to higher powers of $\nu$ and $\bar\nu$). One can use much of the technology developed in the SYK context, though unlike previous generalisations \cite{Gu:2016oyy,Gross:2016kjj,Berkooz:2016cvq}, the couplings are not taken to be spatially disordered. Some further work on SYK can be seen in \cite{You:2016ldz,Anninos:2016szt,Jevicki:2016bwu,Bagrets:2016cdf,Jevicki:2016ito,Banerjee:2016ncu,Garcia-Garcia:2016mno,Fu:2016vas,Witten:2016iux,Cotler:2016fpe,Klebanov:2016xxf,Blake:2016jnn,Davison:2016ngz,Peng:2016mxj,Liu:2016rdi,Krishnan:2016bvg,Magan:2016ehs,Ferrari:2017ryl,Garcia-Garcia:2017pzl,Li:2017hdt,Gurau:2017xhf,Mandal:2017thl}.

The model is also an example of a conformal field theory -- in this case free fermions -- which is perturbed by a large set of marginal operators with random coefficients, some of them being relevant and some irrelevant. In the usual lore, couplings along marginally relevant directions increase, and couplings along marginally irrelevant directions decrease, as one flows towards the IR. In this case, however, RG mixes the two and the anomalous dimension of each operator depends on all of the other random couplings. If we fix the scale at which we define the theory, one expects it will eventually flow in the direction of some relevant operators, but the details and rates at which this happens are unclear. 

It is also interesting to ask whether the theory is renormalizable, since we are necessarily turning on marginally irrelevant perturbations -- but on the other hand they are in the same statistical ensemble as the marginally relevant ones. For the model in \eqref{IntroMod}, the two point function cannot be solved completely, but we compute it at large distances (in the sense that we make precise below) for all values of $J$, without exhibiting a singularity, or the generation of a mass gap, in the physical domain\footnote{These statements are true in the large N limit, with a significant caveat that we have so far solved only for the two point function.}. From it we can extract an effective $\beta$ function 
\begin{equation}
\beta(J)=4\pi^2 J^3
\label{intro-beta}
\end{equation}
and study the above questions. The $\beta$ function indicates that the interaction is marginally irrelevant so the theory is not conformal. It will also turn out not to be conformal in the limit of $J\to\infty$. Rather, the theory requires regularization for any value of $J$, and this regularization breaks conformal invariance already at the level of the 2-pt function.

Useful reference points are the Thirring or Gross-Neveu models \cite{Moshe:2003xn,ZinnJustin:2002ru,Gross:1974jv} (which are equivalent in our case), where the interaction term is non-random and diagonal between the left and right moving fermions. The $\beta$ function of these models is well understood -- in 2D, the leading term is quadratic in the 4-fermion coupling, and the interaction is marginally relevant for one sign of it. The theory then develops a mass gap (which is seen in \cite{Gross:1974jv} in perturbation theory via the appearance of a tachyonic pole at very low energies). 

At finite $N$ we can expect a similar behavior in the extreme IR for each realisation, since each contains both relevant and irrelevant operators, and the former will eventually grow to be large, most likely triggering a mass gap. This, however, is the equivalent of the statement that in the 0+1 SYK the very late time behavior, or the behavior of states close enough to the ground state, differ from realisation to realisation \cite{Cotler:2016fpe}\footnote{See \cite{Balasubramanian:2014gla} for earlier work on black holes and random matrix theory.}. 

The large $N$ ensemble average does not capture this expectation, but it captures the flow prior to chiral $\SO(N)_L\times \SO(N)_R$ symmetry breaking, where it is modified relative to the Thirring and Gross-Neveu models -- in our models we see no pole in the physical regime. Furthermore, the leading quadratic term in the $\beta$ function, which drove the RG in the Thirring and Gross-Neveu models, vanishes for the ensemble (in the large N limit). The $\beta$ function \eqref{intro-beta} is positive for any value of $J$, and it is driven by the cubic terms in the random couplings of the $\beta$ function. The latter is non-universal in the Thirring or Gross-Neveu models, but it is universal in our case for schemes preserving the statistical chiral symmetry. In fact, we reproduce the result \eqref{intro-beta} from suitable average considerations of the perturbative conformal field theory results reported in \cite{Gaberdiel:2008fn,Behr:2013vta}. 

The 2-pt function therefore does not show signs of the chiral phase transition, at leading order in $1/N$, for any value of the coupling. It is known that in low dimensions quenched disorder can smooth out thermal phase transitions \cite{Imry:1975zz}. Our results generalize that statement to translationally invariant theories, with the chiral phase transition being smoothed out at this order. To our knowledge this is the first demonstration of smoothing out of a phase transition by translationally invariant ensemble average in a field theory.

One final note about the application of this model to black hole physics. The obvious analogy is to consider this model as related to a black hole in $AdS_3$. However, we do not want to consider this as the full dual to the entire $AdS_3$ space because the $\beta$ function is positive and the theory will not be asymptote to empty $AdS_3$. Perhaps a better way to think about this model is as describing the effective interactions of degrees of freedom inside a BTZ black hole \cite{Berkooz:2016cvq}. One then needs to 1) "prevent" these degrees of freedom from appearing at high momenta, and 2) couple additional "probe" fields to the model, which will mimic the degrees of freedom outside the black hole\footnote{A related statistical mechanical model appears in \cite{Banerjee:2016ncu}.}. 

This paper is organised as follows. In section \ref{defmodel}, we define the Random Thirring model. In section \ref{SDsec}, we compute the fermion 2-pt function by solving the Schwinger-Dyson (SD) equations in the IR, for all values of $J$, and show how the 2-pt function flows. We also discuss the reparameterization invariance of the model. In section \ref{pCFT}, we compare our model to the flow of the usual Thirring model (or the Gross-Neveu model in our case). We compute the $\beta$ function and show which terms are washed out due to the randomness of the coupling, and which remain active. Section \ref{disc} contains some concluding remarks.

During the completion of this work, the preprint \cite{Turiaci:2017zwd} appeared which contains a 1+1 generalisation of the SYK model with the same interaction term. Although some of the intermediate steps are similar, the UV starting point of our model is different than the one described there, as is our solution and its implications for the RG flow of the model.

\vspace{20pt}
\section{The Random Thirring, or SYK, model in 1+1 dimensions}
\label{defmodel}

Consider a 2d non-chiral theory with N left moving $\nu^i$ and N right moving ${\bar\nu}^i$ ($i=1,\dots ,N$). Its free theory action and partition function in Euclidean space are given by 
\begin{equation}
\label{FreeMaj2}
S_{\text{free}}(\nu,\,\bar{\nu}) = \frac{1}{2\pi} \int d^2z \ \sum_i\left[ {\bar\nu}^i \partial_z {\bar\nu}^i  + \nu^i  \partial_{\bar z}  \nu^i  \right],\quad Z=\int D\nu D{\bar \nu} \,e^{-S(\nu,\,\bar{\nu})}\,.
\end{equation}
The free propagators are
\begin{equation}
\begin{aligned}
\langle \nu_i(z_1,\bar z_1) \nu_j(z_2,\bar z_2) \rangle = \frac{\delta_{ij}}{z_1 - z_2} \equiv G_{\nu}^0(\vec x_1,\vec x_2) \delta_{ij}, \\ 
\langle \bar \nu_i(z_1,\bar z_1) \bar \nu_j(z_2,\bar z_2) \rangle = {\delta_{ij} \over \bar z_1 - \bar z_2}\equiv G_{\bar \nu}^0(\vec x_1,\vec x_2) \delta_{ij} 
\end{aligned}
\end{equation}
where $\vec x$ denotes the pair $z,{\bar z}$. Their Fourier transforms equal
\begin{equation}
  G^{0}_\nu(p)=\frac{i\pi}{\bar p}, \quad G^{0}_{\bar\nu}(p)=\frac{i\pi}{p} 
\end{equation}
For more details on our conventions, see appendix \ref{2dcft}.

The free theory is $\SO(N)_L \times \SO(N)_R$ invariant, with conserved currents
\begin{equation}
J^{[ij]}(z) = i \ \frac{\nu^{i}\nu^{j}(z) - \nu^{j}\nu^{i}(z)}{2} ,\quad {\bar J}^{[ij]}(\bar z) =i \ \frac{{\bar\nu}^{i}{\bar\nu}^{j}({\bar z}) - {\bar\nu}^{j}{\bar\nu}^{i}({\bar z}) }{2}  .
\end{equation}
defined for a pair $[ij]$ with $i<j$. Using these currents, we can add an interaction of the form
\begin{equation}
{\cal L}_{int}=- \sum_{i<j,k<l}J_{ij;kl} J^{[ij]}{\bar J}^{[kl]}={1\over 4}\sum_{i,j,k,l}J_{ij;kl}\nu^{i}\nu^{j}{\bar\nu}^{k}{\bar\nu}^{l},\ \ \ J_{ij;kl}=-J_{ji;kl}=-J_{ij;lk}
\label{model}
\end{equation}
The model we discuss in this work consists in treating the coupling constants $J_{ij;kl}$ as random variables drawn independently from a gaussian distribution defined by
\begin{equation}\label{NewJAv}
\langle \langle J_{ij;kl}^2 \rangle \rangle = \frac{2J^2}{ N^3}
\end{equation} 
where $\langle \langle\ \  \rangle \rangle$ denotes ensemble average over $J_{ij;kl}$.

The resulting model can either be viewed as a translationally invariant continuum generalisation of the SYK model in 1+1 dimensions, or as a random cousin of the massless $\SO(N)$ Thirring model, where by the latter we mean the theory whose interaction Lagrangian involves a fixed diagonal coupling $J_{ij;kl} \sim g\,\delta_{ki}\delta_{lj}$ 
\begin{equation}
\label{ThirSec2}
  {\cal L}_{int} \propto \frac{g}{2} \sum_{i,j} J^{[ij]}\,{\bar J}^{[ij]}\,.
\end{equation}
We will refer to the model in equation \eqref{FreeMaj2} and \eqref{model} as the random Thirring model. 

Though much of our discussion will be about \eqref{model}, there are two natural generalisations to consider. The first involves a higher order interaction term,
\begin{equation}
{\cal L}_{q,int}=\sum_{i_1,..i_q,k_1,..k_q}J_{i_1..i_q;k_1..k_q} \nu^{i_1}...\nu^{i_q}{\bar\nu}^{k_1}...{\bar\nu}^{k_q},\ \ q>2
\label{2qmodel}
\end{equation}
as already discussed in \cite{Turiaci:2017zwd}. The second involves the inclusion of a low-pass filter multiplying each fermion in Fourier space
\begin{multline}
{\cal L}_{q,int}=\int dp_1..dp_q du_1..du_q \delta(\sum p_i+\sum u_i) \sum_{i_1,..i_q,k_1,..k_q}J_{i_1..i_q;k_1..k_q} \\
(F(p_1)\nu^{i_1}(p_1))...(F(p_q)\nu^{i_q}(p_q))
(F(u_1){\bar\nu}^{k_1}(u_1))...(F(u_q){\bar\nu}^{k_q}(u_q)),\ \ q\ge 2,
\end{multline}
where the filter $F(k)$ decays at large momenta. The use of such filters, introduced in this context in \cite{Berkooz:2016cvq}, might be useful for holography in order to force a distinction between UV and IR degrees of freedom, but clearly it will be very restricted from a local field theory point of view. 

Our model \eqref{model} has two types of symmetries : exact and statistical. The first ones hold for any realisation of the couplings, whereas the second ones only arise after carrying the coupling average. Poincar\'e symmetry and a $\mathbb{Z}_2$ symmetry $\nu\rightarrow -\nu$ and ${\bar\nu}\rightarrow{\bar \nu}$ for even q, or its composition for odd q, are exact symmetries of the Lagrangian. Once the ensemble average is carried out, and assuming self-averaging, i.e. that any realisation behaves similarly to the ensemble average, there is an emergent $\SO(N)_L \times \SO(N)_R$ symmetry, rotating left and right Majorana fermions independently, together with a parity exchanging $\nu^i\leftrightarrow {\bar\nu}^i$.

As in the case of the SYK model, we will sum over different realisations of the couplings $J_{ij;kl}$. If the model is self-averaging then this should reproduce, up to small corrections, the results of any specific realisation. However, for a specific realisation the random Thirring model and its $q>2$ generalisations are ordinary field theories. For $q>2$ the interaction term is irrelevant and we don't expect it to generate new interesting dynamics in the IR (although it can generate one in some intermediate scale). For $q=2$ the situation is more interesting, as we perturb the theory by a large number of operators, some of them relevant and some others irrelevant. In fact all the couplings mix together under RG flow but the qualitative picture that some are (marginally) relevant and some are (marginally) irrelevant is expected to hold. Further discussion on these points will appear in section 4.

\vspace{20pt}
\section{Two Point Function : Schwinger-Dyson Equations}
\label{SDsec}

To derive the Schwinger-Dyson (SD) equations for our models, we use the standard replica method \cite{Sachdev:2015efa} to perform the disorder average. If we further assume the replica symmetry is unbroken, the single replica action acquires the following bi-local interaction term
\begin{equation}\label{BilocalIntineta}
  S_{int} \propto J^2 N  \int d^2\!  \vec x_1 d^2\!  \vec x_2 \left( {\sum_i \bar{\nu}^i(\vec x_1) \, \bar\nu^i(\vec x_2 )\over N} \right)^2 \left( {\sum_j\nu^j (\vec x_1 ) \,\nu^j(\vec x_2 ) \over N} \right)^2\,.
\end{equation}

Next, we introduce two Lagrange multipliers $\Sigma_\nu(\vec x_1,\vec x_2), \Sigma_{\bar \nu}(\vec x_1,\vec x_2)$, whose equations of motion impose the constraints
\begin{equation}
G_\nu(\vec x_1,\vec x_2) =\frac{1}{N} \sum_i \nu^i(\vec x_1) \nu^i(\vec x_2)\,, \quad  G_{\bar \nu}(\vec x_1,\vec x_2) =\frac{1}{N}\sum_i \bar \nu^i(\vec x_1) \bar \nu^i(\vec x_2)\,.
\end{equation}
This is achieved by adding to the action the extra term
\begin{equation}
\delta S \propto N\int d^2\!  \vec x_1 \int d^2 \! \vec x_2 \left[ \Sigma_\nu( \vec x_1,\vec x_2) \left( G_\nu(\vec x_1,\vec x_2) - {\sum_i \nu^i(\vec x_1) \nu^i(\vec x_2) \over N}  \right) + (\nu^i \to \bar \nu^i)  \right] 
\end{equation}
with $\Sigma(\vec x_1,\vec x_2)=-\Sigma(\vec{x}_2,\vec{x}_1)$. The resulting action is quadratic in the fermions. Performing their gaussian integral, we are left with the action:
\begin{equation}
\label{ColAction}
\begin{split}
- \frac{S}{N} & \propto   \log \text{Pf} (\partial_{\bar z} -\Sigma_\nu(\vec x_1, \vec x_2)) + \log \text{Pf} (\partial_{  z} -\Sigma_{\bar \nu}(\vec x_1, \vec x_2))  - \int d^2\! \vec x_1 d^2\! \vec x_2  \\
& \ \ \ \ \times \left[ \Sigma_\nu( \vec x_1,\vec x_2)   G_\nu(\vec x_1,\vec x_2) + \Sigma_{\bar \nu}( \vec x_1,\vec x_2)  G_{\bar \nu}(\vec x_1,\vec x_2) - \frac{J^2}{2} G_\nu(\vec{x}_1,\vec{x}_2)^2 G_{\bar \nu}(\vec{x}_1,\vec{x}_2)^2     \right]
\end{split}
\end{equation}
The equations of motion for $G_\nu(\vec{x}_1,\vec{x}_2)$ and $G_{\bar \nu}(\vec{x}_1,\vec{x}_2)$ are 
\begin{equation}\label{DefG}
\begin{split} 
\Sigma_\nu(\vec x_1,\vec x_2 )   &=  J^2 G_\nu(\vec x_1,\vec x_2 )\, \bar{G}_{\bar \nu}(\vec x_1,\vec x_2 )^2\,, \\
\Sigma_{\bar \nu}(\vec x_1,\vec x_2 )   &=   J^2  \bar{G}_{\bar \nu}(\vec x_1,\vec x_2 )  \, G_\nu(\vec x_1,\vec x_2 )^2\,,
\end{split}
\end{equation}
whereas the equations of motion for $\Sigma_\nu(\vec{x}_1,\vec{x}_2)$ and $\Sigma_{\bar \nu}(\vec{x}_1,\vec{x}_2)$ are then:
\begin{equation}\label{DefSigma}
\begin{split}
G_\nu(\vec p_1,\vec p_2 )^{-1} &=  G^{0}_{\nu}(\vec p_1,\vec p_2)^{-1} - \Sigma_\nu(\vec p_1,\vec p_2)\,,  \\
G_{\bar \nu}(\vec p_1,\vec p_2 )^{-1} &=  \bar{G}^{0}_{\bar \nu}(\vec p_1,\vec p_2)^{-1} - \bar{\Sigma}_{\bar \nu}(\vec p_1,\vec p_2)\,. 
\end{split}
\end{equation}

The same procedure can be carried out for the 2q fermion interaction model \eqref{2qmodel}. The corresponding SD equations are
\begin{equation}\label{FullSDEqnq1}
\begin{aligned}
 G_\nu(\vec p_1,\vec p_2 )^{-1} &=  {G^{0}_\nu}(\vec p_1,\vec p_2 )^{-1} -    \Sigma_\nu(\vec p_1,\vec p_2 )\,,      \\
 G_{\bar \nu}(\vec p_1,\vec p_2 )^{-1} &=   G^{0}_{\bar \nu}(\vec p_1,\vec p_2 ){-1}   -    \Sigma_{\bar \nu}(\vec p_1,\vec p_2 )\,,\\
  \Sigma_\nu(\vec x_1,\vec x_2 )  &=  J^2\,G_\nu(\vec x_1,\vec x_2 )^{q-1}\, G_{\bar \nu}(\vec x_1,\vec x_2 )^{q}\,,  \\
  \Sigma_{\bar \nu}(\vec x_1,\vec x_2 ) & = J^2\, G_{\bar \nu}(\vec x_1,\vec x_2 )^{q-1}\, G_\nu(\vec x_1,\vec x_2 )^{q}\,.
\end{aligned}
\end{equation}  

\vspace{10pt}
\subsection{A comment on reparametrization invariance}\label{sec:reparam}

Consider the $J\to \infty$ limit of the collective field action \eqref{ColAction} for the 2q fermion model \eqref{2qmodel}. This results in dropping the dependence on the free propagator in the equations of motion \eqref{FullSDEqnq1}. It is convenient to rewrite the first two equations in real space
\begin{equation}
\label{SDrspace}
\begin{aligned}
  \int d^2z\, G_\nu(z',z;\bar z',\bar z) \Sigma_\nu(z, z'';\bar z , \bar z'') &= -\delta(z'-z'') \delta(\bar z'-\bar z'')\,, \\
  \int d^2z\, G_{\bar \nu}(z',z;\bar z',\bar z) \Sigma_{\bar \nu}(z, z'';\bar z , \bar z'') &= -\delta(z'-z'') \delta(\bar z'-\bar z'')\,,
\end{aligned}
\end{equation}
to discuss their symmetries. As pointed out in \cite{Turiaci:2017zwd}, these equations of motion appear to be invariant under reparametrization. 

For the case $q=2$, which we will argue below is the only physically interesting one for us, we can also keep the free kinetic term and reintroduce it into equations \eqref{SDrspace}. In this case the equations are invariant under conformal transformation $z\to f(z)$ and $\bar z \to \bar f(\bar z)$. To see this, consider the 2-pt function transformations
\begin{equation}\label{eq:TwoPointTransf}
\begin{aligned}
  G_\nu(z,z';\bar z,\bar z') &\to [f'(z)\,f'(z')]^{\Delta_L}  [\bar f'(\bar z)\bar f'(\bar z')]^{\bar\Delta_L}\, G_\nu(f(z),f(z');\bar f(\bar z),\bar f(\bar z'))\,, \\
  G_{\bar \nu}(z,z';\bar z,\bar z') &\to [f'(z) f'(z')]^{\Delta_R}  [\bar f' (\bar z)\bar f'(\bar z')]^{\bar\Delta_R} G_{\bar \nu}(f(z),f(z');\bar f(\bar z),\bar f(\bar z'))\,.
\end{aligned}
\end{equation}
The last two equations in \eqref{FullSDEqnq1} determine the transformation for the self-energies $\Sigma_\nu$ and $\Sigma_{\bar{\nu}}$, and the invariance of equations \eqref{SDrspace} (with the free kinetic term reinstated) requires that  $(\Delta_L,\Delta_R)=({1 \over 2},0)$ and $({\bar\Delta}_L,{\bar\Delta}_R)=(0,{1 \over 2})$
(for a general $q$, dropping the kinetic term, invariance under conformal transformations implies only that  $\Delta_L + \Delta_R = \bar{\Delta}_L + \bar{\Delta}_R = \frac{1}{q}$).

The conclusion from this analysis would suggest to solve the SD equations with a scale invariant ansatz for any $J$ for $q=2$ (and for large $J$ for $q>2$). Below, we will argue this is not necessarily the case. This quantum field theory requires regularization, which breaks conformal invariance. A-priori this symmetry may or may not be restored in the IR (or may be restored upon fine tuning some operators away). Actually, here it will not be restored. We will see this explicitly for the $q=2$ theory - the solution would not be scale invariant (although in a mild sense). It is important to emphasize that the regulator needs to be included already in the action, and the breaking of scale invariance is explicit (all the way to the IR) and not spontaneous. 

As an aside we would also like to comment that the conformal symmetry of the action is also broken in the IR, beyond the effects of the regulator. The reason is that the action \eqref{ColAction} should be appended by boundary conditions, or, in the language of the equations of motions, the solutions are legitimate only when the functions decay fast enough at infinity. This is nothing but the usual argument why ordinary 2D CFTs can have only an $\SL(2)\times \SL(2)$ symmetry on the plane - all other generators of the Virasoro algebra introduce singularities at infinity, which change an n-point function into an n-point function in the presence of an operator at infinity. 

This situation is different from reparametrization invariance in the SYK model in 0+1 dimensions. In that case, at the level of the 2-pt function, the action was consistently reparametrization invariant and a scale invariant solution could be found. Reparametrization was then broken spontaneously. This pattern of symmetry breaking was used heavily in computing the 4-pt function, although in order to make sense of it one had to eventually re-introduce the explicit breaking of reparametrization. Here the cut-off breaks reparametrization explicitly and the solution is not scale invariance. The net result of an explicit breaking of scale invariant (let alone reparametrization) is similar here and in the 0+1 SYK model, but the differences in the intermediate stage suggests that computing the 4-pt function would be considerably more involved.

\vspace{10pt}
\subsection{Solutions of the SD equations -- general q}

In this subsection we will look for solutions to the SD equations \eqref{FullSDEqnq1} for various values of $q$. We will show that for $q>2$ the interaction term is subleading in the IR, and for $q=2$ it introduces an unusual form of a logarithmic running into the 2-pt function. We will assume translation invariance and, to evaluate the importance of various terms, we will carry out a scaling argument in the IR. We will also assume that Lorentz symmetry (Euclidean rotation) is unbroken throughout the flow and hence the UV Lorentz numbers of the fermions remain fixed. We will also assume that parity is unbroken. Under these assumptions
\begin{equation}
G(z_1-z_2,\bar z_1-\bar z_2) \equiv G_{\nu}(z_1,\bar z_1, z_2, \bar z_2) = G_{\bar \nu}(\bar z_1,  z_1, \bar z_2,  z_2) 
\end{equation}
and one can similarly define $\Sigma(z,\bar z)$. Then the SD equations \eqref{FullSDEqnq1} collapse to 
\begin{eqnarray}\label{SimpSD}
G(p, \bar p)^{-1} = G^{(0)}(p, \bar p)^{-1} - \Sigma(p,\bar p)\,, \quad \Sigma(z,{\bar z}) = J^2 G(z,{\bar z})^{q-1} G({\bar z},z)^{q}
\end{eqnarray}

Consider a scaling solution
\begin{equation}
G(\lambda z,{\bar \lambda} {\bar z})=\lambda^{-2\Delta_L}{\bar\lambda}^{-2\Delta_R}G(z, {\bar z}),\quad G(\lambda^{-1}\,p, \bar{\lambda}^{-1}\bar p)=\lambda^{-2\Delta_L+1}{\bar\lambda}^{-2\Delta_R+1}G(p,\bar p)
\label{scaling}
\end{equation}
with $2(\Delta_L-\Delta_R)=1$, as dictated by invariance under Euclidean rotations. These solve the SD equations if
\begin{equation}
\Delta_L + \Delta_R = \frac{1}{q}  \implies \quad  2\Delta_{L} =  {1 \over q} + {1 \over 2} \quad , \quad 2\Delta_R  = {1 \over q} -{1 \over 2}
\end{equation}
We treat the $q>2$ and $q=2$ cases separately.

For $q>2$ we obtain $\Delta_R < 0$, which means the solutions are not physical. If the model is self-averaging (which we assumed before) then correlators should be close to being correlators in a unitary field theory. This necessitates that both $\Delta_L$ and $\Delta_R$ are non-negative. Or, related to it, the free propagator $G^{(0)}(p_z,p_{\bar z}) \sim {1 \over \bar p}$ already scales as $\bar \lambda$ and is leading at low momenta. Basically this means the theory is free in the IR\footnote{One might still be able to work in intermediate energy regimes, where the above approximation might be correct, but it will break down at sufficiently low energies.}. This is what one expects from field theory considerations as well. Deforming a free fermion theory by operators of the form $\nu^q{\bar \nu}^q$, which are operators of dimension $(q/2,q/2)$, the perturbation is irrelevant\footnote{This is also potentially where we differ from \cite{Turiaci:2017zwd} which add the interaction term to an altogether different theory.}. Thus, we will not consider the $q>2$ case further. 

\vspace{10pt}
\subsection{Solutions of the SD equations for q=2}

The scaling argument \eqref{scaling} suggests that the solutions to the SD equation \eqref{SimpSD} for the $q=2$ model are 
\begin{equation}
G(z,\bar z) = \frac{C}{z}, \ \ \ \Sigma(z,\bar z) = \frac{C^3 J^2}{z \bar z^2}\,,
\end{equation}
for some constant $C$. The difficulty with this ansatz originates in the Fourier transform of $\Sigma(z,\bar z)$. The latter is not well defined and requires the introduction of a scale. Following the conventions in appendix \ref{2dcft}, it is given by
\begin{equation}
\label{FT1byzbz2}
  \frac{1}{z \bar z^2} \propto \int {d^2\!p_z \over (2\pi)^2} e^{-i (pz + {\bar p} \bar z)}  {\bar p} \log(\Lambda^2/|p|^2) 
\end{equation}
We would therefore like to see what the solution to the SD equations is. Our approach will be to modify our ansatz to include some dependence on logarithms, which will turn out to be essential to encode the RG flow of the model. We will present the large J solution in the IR before proceeding to the all-J solution (still at large distances).  

\vspace{10pt}
\subsubsection{The Large J IR solution}

We first consider the large $J$ limit, where the free propagator terms in the SD equations can be neglected. If we interpret the presence of the logarithms above as inducing some soft RG flow on top of some leading power law term, as dictated by the scaling argument \eqref{scaling}, it may appear natural to consider the ansatz
 \begin{equation}\label{Jansatz}
G(z,\bar z) = \frac{A}{z} \left(\log(z{\bar z}\Lambda^2)\right)^\alpha\,, \quad \Sigma(z,\bar z)=\frac{A^3J^2}{z{\bar z}^2} \left(\log(z{\bar z}\Lambda^2)\right)^{3\alpha}\,.
\end{equation}
This depends on two parameters $A$ and $\alpha$, to be determined by solving the SD equations, and a cut-off scale $\Lambda$, whose role as a UV cut-off will become more apparent below. In this ansatz, we solve the "interaction" SD equation exactly, and then we analyse whether there is any regime of low momentum where our ansatz is self-consistent, i.e., the "propagator" SD equation is solved (at least in some approximation). 

Working in momentum space, we observe that the Fourier transforms $G(p,{\bar p})$ and $\Sigma(p,{\bar p})$ of our ansatz \eqref{Jansatz} satisfy
\begin{equation}
\begin{aligned}
G(p,{\bar p}) &=\partial_{\bar p} {\hat G}(p,{\bar p})\,, \quad \text{with} \quad {\hat G}(p,{\bar p}) \equiv {A \over i} \int {d^2\!z \over 2 } {1 \over z{\bar z}} \log^\alpha(\Lambda^2z{\bar z})\,e^{ipz+i{\bar p}{\bar z}} \\
\partial_{\bar p}\Sigma(p,{\bar p}) &= A^3J^2i \int {d^2\!z \over 2} { \log^{3\alpha}(\Lambda^2z{\bar z}) \over z{\bar z}}\,  e^{ipz+i{\bar p}{\bar z}}\,.
\end{aligned}
\end{equation}
Hence, the integral we need to evaluate is
\begin{equation}
\label{original}
\begin{aligned}
F_\alpha(|p|/\Lambda)&\equiv \int {d^2\!z \over 2} {1\over z{\bar z}} \log(\Lambda^2z{\bar z})^\alpha e^{ipz+i{\bar p}{\bar z}}= \int { d^2\!\vec x   \over |\vec x|^2 } \log^\alpha(\Lambda^2 |\vec x|^2)\, e^{i \vec p . \vec x} \\
 &= 2^\alpha \int { dr \over r } \log^\alpha (\Lambda r) \int d\theta e^{i p r \cos \theta} = 
(2 \pi) \ 2^\alpha   \int_0^\infty {dr \over r} \log^\alpha(\Lambda r) J_0(|p| r)\,,
\end{aligned}
\end{equation}
where we used the integral representation of the Bessel function $J_0(x)$ to perform the integration over the angular variable and $|p|^2=4p\,\bar p={\vec p}^2$ according to our conventions in appendix \ref{2dcft}.

To evaluate this integral, we start from the convergent formula\footnote{See formula 6.771 in \cite{gradshteyn}, for example.}
\begin{equation}
\begin{aligned}
G_\epsilon(|p|/\Lambda) &\equiv \int_0^\infty dr \ (\Lambda r)^\epsilon \  {\log(\Lambda r) \over r} \ J_0(|p| r) \\
&= {\Gamma({\epsilon \over 2}) |\Lambda/p|^\epsilon \over \Gamma(1-{\epsilon \over 2}) 2^{2 -\epsilon}  } \left[ \psi({\epsilon \over 2}) + \psi(1-{\epsilon \over 2}) +2 \log(2 |\Lambda/p| )\right]\,,
\end{aligned}
\end{equation}
valid for $0<\epsilon< \frac{3}{2}$ and where $\psi(x)$ is the Digamma function. Notice we can compute $F_\alpha(|p|/\Lambda)$ for $\alpha = n \in \mathbb Z$ from
\begin{equation}
  F_{n+1}(|p|/\Lambda)= 2^{n+2}  \pi \lim_{\epsilon \to 0} \partial_\epsilon^n G_\epsilon(|p|/\Lambda)\,.
\label{Fdev}
\end{equation}

To proceed, we can write $G_\epsilon(|p|/\Lambda)$ for small $\epsilon$ as
\begin{equation}
\begin{split}
G_\epsilon(|p|/\Lambda) =&   \left( {1\over\epsilon } + (\log 2- \gamma)+{\cal O}(\epsilon) \right)   e^{\epsilon \log(\Lambda/|p|)} \biggr(-{1\over \epsilon} -  \gamma +{\cal O}(\epsilon) + \log(2\Lambda/|p|) \biggr) \\
=& - {1 \over \epsilon^2} +  \sum_{n=0}^\infty  {\epsilon^n \log^{n+2}(\Lambda/|p|) \over n+2} \left[ 1 + {\cal O}({1 \over \log(\Lambda/|p|)}) \right] \\
\end{split}
\end{equation}
where $\gamma$ is the Euler's constan.
Plugging this leading logarithm into \eqref{Fdev}, and dropping the divergent momentum independent $\epsilon^{-2}$ term and the subleading momentum dependent logarithms, we get
\begin{equation}
\label{GenFormFn}
  F_n(|p|/\Lambda) = \frac{\pi}{n+1}  \log^{n+1}( \frac{\Lambda^2}{|p|^2})  \left( 1 +{\cal O}({1 \over \log(\Lambda/|p|)})\right) \,.
\end{equation}

Removing the $1/\epsilon^2$ divergences in the procedure above resembles dimensional regularization, but since we do not understand the renormalization of the model well enough we provide an explicit argument. In appendix \ref{alter}, we provide a specific regularisation of our original integral \eqref{original} leading to the same conclusion in the same IR regime. The gist of the argument is that since $\Lambda$ is our cut-off, we should not do the integral below a distance $\Lambda^{-1}$. Since the integral is convergent for $\epsilon > 0$, we can write
\begin{equation}
\int_{\Lambda^{-1}}^\infty dr\dots = \int_0^\infty dr \dots -  \int_0^{\Lambda^{-1}} dr \dots
\end{equation}
Changing variables to $y=\Lambda r$, working in the limit of small $|p|\Lambda$ and expanding $J_0$ (since the range of integration is finite in $y$), we obtain in the second integral a divergent $1/\epsilon^2$ piece which cancels the divergence from the first integral, justifying the derivation of equation \eqref{GenFormFn}.

We would like to extend our conclusion for integer $\alpha$ to a general one 
\begin{equation}
F_\alpha(|p|/\Lambda)=\frac{\pi}{\alpha+1} \left(\log (\Lambda^2/|p|^2)\right)^{\alpha+1} \left(1+{\cal O}(1/\log(|p|/\Lambda)) \right)\,.
\label{Fresult}
\end{equation}
by the following qualitative argument (which becomes more precise for $\alpha$'s which are negative enough). Working with the dimensionless variable $y=\Lambda\,r$,
\begin{equation}
F_\alpha(|p|/\Lambda) = \pi\, 2^{\alpha+1}\,\int_1^\infty \frac{dy}{y} \left(\log y\right)^\alpha J_0(|p|\,y/\Lambda)\,.
\label{Fintegral}
\end{equation}
When $|p|/\Lambda$ is small, we can split the integral \eqref{Fintegral} into $[1,\, C\, \Lambda/|p|]$ and $[C\,\Lambda/|p|,\,\infty)$, where $C$ is finite and smaller than 1. In the first interval, the argument of the Bessel function is generically small and the function can be approximated by one. The resulting integral immediately reproduces \eqref{Fresult}. We are left to argue the contribution from the second interval is subleading. This is so for two reasons. First, when the argument $z$ of the Bessel function is large, the latter can be approximated by $z^{-1/2}\,\cos (z - \pi/4)$ (up to coefficients of order one). The integral involves a non-oscillating function multiplied by this oscillating decaying function. Second, for $\alpha$ negative and large enough in absolute value, the logarithm is small in this range.

In any case, the integral can also be evaluated numerically for the specific values of $\alpha$ that interest us in the same regime
$|p|/\Lambda \ll 1$. These confirm the expression \eqref{Fresult}. In the next subsection we present additional arguments for the validity of this large J solution.

Using \eqref{Fresult} determines the two Fourier transforms
\begin{equation}
\begin{aligned}
  G(p,{\bar p}) &=  i\frac{A\pi}{\bar p} \left(\log \frac{\Lambda^2}{|p|^2}\right)^{\alpha}\,, \\
  \partial_{\bar p}\Sigma(p,{\bar p}) &= i\frac{A^3J^2\pi}{3\alpha+1}\left(\log \frac{\Lambda^2}{|p|^2}\right)^{3\alpha+1} \quad \Rightarrow \quad \Sigma(p,{\bar p}) = i{\bar p} \frac{A^3J^2\pi}{3\alpha+1}\left(\log \frac{\Lambda^2}{|p|^2}\right)^{3\alpha+1}\,.
\end{aligned}
\end{equation}
The SD equation $G(p,{\bar p})\,\Sigma(p,{\bar p}) = -1$ requires
\begin{equation}
4\alpha+1=0,\ \ \ -\frac{A^4J^2 \pi^2}{3\alpha+1}=-1 \quad \Rightarrow \quad \alpha = -\frac{1}{4}\,, \,\,\, A^4=\frac{1}{4\pi^2\,J^2}\,.
\end{equation}

As a quick consistency check that this is a sensible result, note that $A$ is positive. This is a necessary condition because it has to correspond to a 2-pt function
\begin{equation}
\sum_i \langle \phi | \nu |i\rangle e^{-\tau E_i} \langle i | \nu | \phi \rangle
\end{equation}
(where $\nu$ is an Hermitian operator) which is positive.

\vspace{10pt}
\subsubsection{Solution for any J}
 
Another regime to consider in coupling space is small $J$, by carrying out perturbation theory in $J$. We can actually do better and find an all-$J$ solution, by remaining in the IR, i.e. large $\Lambda^2 z{\bar z}$ or $\Lambda^2/|p|^2$. We will assume an ansatz of the form  
\begin{eqnarray}\label{perturbative anzatz}
G(z,\bar z) &=& \frac{1}{z}  \sum_{n=0}^\infty  a_n J^{2n} \log^n(\Lambda^2 z \bar z) \xrightarrow{|z| \gg \Lambda^{-1}, |p| \ll \Lambda} G(p , \bar p ) = \frac{i \pi}{\bar p}  \sum_{n=0}^\infty  a_n J^{2n} \log^n\frac{\Lambda^2}{|p|^2} \\
\nonumber
\Sigma(z,\bar z) &=& \frac{J^2}{z \bar z^2}  \sum_{n=0}^\infty  b_n J^{2n} \log^n(\Lambda^2 z \bar z) \xrightarrow{|z| \gg \Lambda^{-1}, |p| \ll \Lambda}  \Sigma(p , \bar p ) = i \pi  {\bar p}   J^2\sum_{n=0}^\infty  \frac{b_n}{n+1}\, J^{2n} \log^{n+1}\frac{\Lambda^2}{|p|^2}
\end{eqnarray} 
with $a_0=1$ to match the free theory propagator. Note that this is not the most general expansion in $J$ and in logarithms that we could consider. In fact if we solve \eqref{DefG}-\eqref{DefSigma} perturbatively in $J$, the coefficient of $J^n$ in \eqref{perturbative anzatz} will have corrections which are lower powers of $\log(\Lambda^2 z \bar z)$ than indicated. However, at lengths much larger than the cutoff scale, these terms are subleading. We will show that the above ansatz is self-consistent, and also solves the SD equations with the right limits $J\to 0$ and large J, i.e. it re-sums the perturbative series allowing to interpolate between both expansions, as long as the IR criteria above is satisfied.
 
Instead of plugging our expansions in the SD equations to determine the coefficients $a_n,\,b_n$ recursively, we notice that the relevant Fourier transforms, summarized in eq. (\ref{Fourier2}-\ref{Fourier3}) (whose operation is denoted by ${\cal F}$ below) can be recast as
\begin{equation}
\begin{aligned}
{\cal F}\left[\frac{1}{z}\,F_G\bigl(\log(\Lambda^2z{\bar z}) \bigr)\right] &= \frac{i\pi}{\bar{p}}F_G\bigl(\log(\Lambda^2/|p|^2)\bigr)\,, \\
{\cal F}\left[\frac{1}{z\,\bar{z}^2}\, F_\Sigma \bigl(\log(\Lambda^2z{\bar z}) \bigr)\right] &= i\pi {\bar p}\,F_{\Sigma,I}\bigl(\log(\Lambda^2/|p|^2)\bigr)\,, \\
F_{\Sigma,I}(x)&\equiv\int^x dy F_\Sigma(y)
\end{aligned}
\end{equation}
Here $F_G,F_\Sigma$ should be thought of as a Taylor expansion in its arguments. This allows us to write our SD equations \eqref{DefG} and \eqref{DefSigma} as
\begin{equation}
F_\Sigma(x)=J^2F_G^3(x)\,, \quad \frac{1}{i\pi F_G(x)}=-i\pi \int_0^x dy F_\Sigma(y)+{1 \over i\pi} \,, 
\end{equation}
where the integration constant is determined by the boundary condition at $J=0$, i.e. the absence of self energy ($\Sigma=0$) when the interaction is turned off ($J=0$). Taking derivatives, we derive the ODE equation
\begin{equation}
  F^\prime_G(x)=-\pi^2J^2 F^5_G(x)\,,\quad F_G(0)=1
\end{equation}
The ODE has a unique solution, giving rise to the 2-pt function
\begin{equation}\label{FullTwoPointFunc}
G(z,{\bar z})=\frac{1}{z}\,\frac{1}{\left(1+4\pi^2J^2\log(\Lambda^2\,z{\bar z})\right)^{1/4}}\,.
\end{equation}
The latter interpolates from small to large J, and has the correct limit at $J=0$ and large $J$.

\vspace{10pt}
\subsection{Beta Function}

Our analysis confirms the existence of an RG flow. We can extract the beta function $\beta (J)$ for the coupling $J$ that controls our random Thirring model from the Callan-Symanzik equation satisfied by the fermion propagator $G(z,{\bar z})$
\begin{equation}
\bigl( \Lambda \frac{\partial}{\partial\Lambda} +\beta(J) \partial_J +2\gamma(J)\bigr) {1\over z (1+ 4\pi^2 J^2\log(\Lambda^2 z{\bar z}))^{1/4}}=0\,,
\end{equation}
whose solution is given by
\begin{equation}\label{betagamma}
\beta(J)=4\pi^2 J^3\,, \quad \gamma(J)=\pi^2J^2\,.
\end{equation}

A positive $\beta$ function tells us the effective coupling becomes smaller in the IR. We are perturbing the action by a large number of marginal operators, which have non trivial 3-pt functions between them. This means that some of the coefficients of these operators will grow and some will decrease as we flow to the IR. In fact, which ones grow and which one decrease change as the couplings themselves evolve. If we think about $J$ as measuring the root mean square of the couplings, then some of them clearly decrease, generating a positive $\beta$ function. Still, the result is a bit peculiar if we want to think about $J\to\infty$, at least at finite $N$, which is not the case here. If we have a theory at finite cut-off and we turn on some relevant and irrelevant operators, then eventually some relevant directions will grow (and the irrelevant ones will decrease). This means that we expect that eventually the $\beta$ function will turn negative - we see no traces of this, presumably because of the large N scaling taken.

In the next section, we will gain a better understanding for the positivity of the beta function \eqref{betagamma} by using conformal perturbation theory around the free field fixed point.

\vspace{20pt}
\section{The $\beta$ function in conformal perturbation theory}
\label{pCFT}

To obtain a better understanding of the RG flow, we compare our theory to the Gross-Neveu \cite{Gross:1974jv} and Thirring models (for reviews, see \cite{ZinnJustin:2002ru, Moshe:2003xn}). These have been studied extensively, also in the context of the large N limit. Furthermore, since our model also deforms a free theory by 4-fermion marginal operators, conformal perturbation theory remains a very useful tool for any specific realisation of the couplings $J_{ij,kl}$, when these are small. The main difference is that our couplings are as random as possible (although they have a large statistical symmetry) and the key issue is the interplay between randomness and conformal perturbation theory.

In this section, we show that even though the $\beta$ function is quadratic in the couplings for a given realisation of the couplings $J_{ij,kl}$, this flow is suppressed at large N for the ensemble average in our models. We will argue that contrary to what happens to most known models, the next order contribution, i.e the cubic coupling term in the $\beta$ function, is universal\footnote{This statement holds for any scheme that preserves $\SO(N)_L\times \SO(N)_R$ invariance.}, drives the RG flow discovered in the previous section and in fact agrees with it for all values of $J$, and not just at weak coupling, in the $|p|/\Lambda \ll 1$ regime.

\vspace{10pt}
\subsection{RG in the Thirring and Gross-Neveu models}

As discussed in section \ref{defmodel}, our model is the "disordered" version of the Thirring model, given in equation \eqref{ThirSec2}. The latter is usually treated by introducing a gauge field $A_\mu^a$ and rewriting the interaction as 
\begin{equation}
  \frac{1}{2} g J_\mu J_\mu\rightarrow \frac{A_\mu^2}{2g} + i\,A_\mu J_\mu\,,
\end{equation}
where we switched to a vector notation for the original currents and suppressed gauge indices \cite{ZinnJustin:2002ru, Moshe:2003xn}. The $A_\mu^2$ coupling is driven $g\to \infty$ in the IR. For the case of a single fermion in the fundamental representation, the theory is driven to a massive theory. If the starting UV field theory is a higher level WZW model, then it is expected\footnote{We are indebted to David Kutasov for a discussion of this  point.} to flow in the IR to a coset theory where we quotient by $\SO(N)$.

In our case, it is actually more useful to compare to the Gross-Neveu model \cite{Gross:1974jv}. For Majorana fermions, the Thirring model can be written as Gross-Neveu model by rewriting
\begin{equation}
\sum_{ab} \delta_{ab}J^a{\bar J}^b\sim \sum_{ij}\nu^i\nu^j{\bar\nu}^i{\bar\nu}^j\sim \sum_{ij}  \bigl(\nu^i{\bar\nu}^i\bigr) \bigl(\nu^j{\bar\nu}^j\bigr)
\end{equation}
Following the notation of Dirac fermions in Minkowski space used in \cite{Gross:1974jv}, this model is
\begin{equation}
  {\cal L} = i {\bar \psi}  {\slashed\partial}  \psi +  {g^2 \over 2 N } ({\bar\psi}\psi)^2\,.
\end{equation} 
It is convenient to introduce an auxiliary bosonic field $\sigma$
\begin{equation}
{\cal L} = i {\bar \psi}  {\slashed\partial}  \psi    - \frac{\sigma^2}{2} - \frac{g}{\sqrt N} {\bar \psi} \psi\, \sigma\,,
\end{equation}
because at large $N$ the fermion propagator is not renormalized and the boson propagator becomes 
\begin{equation}
D_R(p,\mu^2)= \frac{-i}{1+\frac{g^2}{2\pi} \log(-\frac{p^2}{\mu^2})}\,.
\end{equation}
One can use the Callan-Symanzik equation to extract the $\beta$ function at large $N$ to be
\begin{equation}
\beta(g) \equiv \mu \frac{\partial g}{\partial \mu} = - \frac{g^2}{2 \pi}
\label{betaGN}
\end{equation}
Hence, if $g>0$, the coupling is marginally relevant and the theory is asymptotically free in the UV. If $g<0$, the coupling is marginally irrelevant and the theory becomes free in the IR. 

We expect a similar structure to exist in the random Thirring model - some of the couplings will be relevant and some irrelevant, their nature changing due to the non-linearity of the $\beta$ functions (as we will see below). In this situation it is more convenient to think about the theory as defined at some cut-off scale which is held fixed, in which case, at least for finite $N$, we expect the theory to emerge along some relevant directions and eventually develop a mass gap. However, this may not be accessible in the large $N$, in the phase where the statistical $\SO(N)_L\times \SO(N)_R$ is preserved. 

For completeness, in the Gross-Neveu model the dynamical scale is given by
\begin{equation}
\Lambda(g) \propto \Lambda\, e^{-{2\pi \over g}}(1+{\cal O}(g)).
\end{equation}
where $\Lambda$ is a UV cutoff used to regularize the theory. In the IR, the theory breaks the chiral symmetry and flows to a massive theory. In fact, the theory is integrable and the complete spectrum of particles, labelled by $n$ (appearing in full multiplets of the symmetry) can be computed exactly and it is given by 
\begin{equation}
m_n=\Lambda(g){N-2\over\pi}\sin	\biggl({n\pi\over (N-2)} \biggr)\,.
\end{equation}

\paragraph{$\beta$ function from conformal perturbation theory.}  

The RG flows for the Thirring and Gross-Neveu models can also be understood using conformal perturbation theory. Consider a 2d CFT (the free fermion theory in our case) deformed by some marginal operators $\mathcal{O}_\alpha$ with conformal dimensions $(h,\bar{h})=(1,1)$
\begin{equation}
S_{\text{CFT}} - \sum_\alpha \lambda_\alpha \int \mathcal{O}_\alpha(z,\bar{z})\,d^2z\,.
\end{equation}
Conformal perturbation theory determines the first contribution to the $\beta$ function of these couplings to be\footnote{We are neglecting numerical factors in this subsection, but we will give precise formulas in subsection \ref{pform}.} \cite{cardy1988conformal}
\begin{equation}\label{BetaConf}
  \beta(\lambda_\alpha) \sim \sum_{\gamma,\sigma} C_{\alpha\gamma\sigma}\lambda_\gamma\lambda_\sigma + \dots
\end{equation}
where $C_{\alpha\gamma\sigma}$ is the 3-pt function $\langle \mathcal{O}_\alpha(z_\alpha,\bar{z}_\alpha) \mathcal{O}_\gamma(z_\gamma,\bar{z}_\gamma) \mathcal{O}_\sigma(z_\sigma,\bar{z}_\sigma)\rangle$ when we normalise the 2-pt functions canonically. 

In our discussion, the deformation is of the form
\begin{equation}\label{SCFTDeform}
S_{\text{CFT}} -  J_{a;{\bar a}} \int J^{a}(z)\,\bar{J}^{{\bar a}}(\bar{z})\,d^2z\,.
\end{equation}
$J^a$ and $\bar J_{\bar a}$ are currents labelled by some index in the adjoint representation of $\SO(N)$ (see appendix \ref{currents} for more details on these conventions). (the same letter $J$ is used to denote both the currents and the coupling constants, but they will be distinguished by their index structure). The operators that we are perturbing by are $\mathcal{O}_{a{\bar a}}(z,\bar z)=J^a(z){\bar J}^{\bar a}(\bar z)$. Their three point functions are of the form
\begin{equation}\label{CintermsofF}
C_{a\bar a,b \bar b, c \bar c} \propto f_{abc}f_{{\bar a}{\bar b}{\bar c}}\,.
\end{equation}
Hence the $\beta$ function \eqref{BetaConf} is
\begin{equation}\label{BetaFunQuad}
  \beta_{aa'} \equiv \beta(J_{a;a'})  \propto \sum_{b,b',c,c'} f_{abc}f_{a'b'c'}\,J_{b;b'}\,J_{c;c'}\,.
\end{equation}
For the Gross-Neveu model,  the couplings equal $J_{a;\bar a} = \delta_{a\bar a} {g \over \sqrt N} $ and $f_{abc}$ are the structure constants of $\SO(N)$ given in appendix \ref{currents}. Using the identity \eqref{soNid}, the beta function \eqref{BetaFunQuad} reduces to 
\begin{equation}
\beta(g) \sim -g^2\,,
\end{equation}
reproducing the behaviour in \eqref{betaGN}.

\vspace{10pt}
\subsection{$\beta$ function in the random Thirring model}

The standard use of conformal perturbation theory is for specific realisations of the set of couplings $J_{a;{\bar a}}$. In the following, we explore its extension to 2d CFT with disordered marginal deformations. 

\paragraph{Flow of ensemble :} The 0+1 SYK model, the random Thirring model, or any of their cousins, have an infinite number of couplings in a given realisation. In principle we should track all of them. Once we declare that couplings are drawn from some class of distributions $f(\{J_{a;{\bar a}}\})$, e.g. Gaussian in our case, the number of parameters is greatly reduced. However, the distribution may flow, and its functional form may change with scale, i.e the ensemble may become a function of scale $f(\{J_{a;{\bar a}}\},\mu)$, in effect introducing more parameters.

There are two equivalent ways of thinking of the RG flow in disordered theories. The first is to think of the flow of ensemble of couplings. In this, the ensemble average is done with a scale dependent distribution denoted by $\langle \langle \rangle  \rangle_\mu$. For example
\begin{equation}\label{eq:flow ensemble}
\langle \langle \prod_{\alpha=1}^n  \lambda_\alpha \rangle  \rangle_\mu  = \int d\lambda_\alpha f(\{\lambda_\alpha\},\mu) \prod_{\alpha=1}^n\lambda_\alpha
\end{equation}
The second is to consider the flow of couplings for a specific realisation drawn from a fixed ensemble. This we denote by the usual $\langle \langle \rangle  \rangle$. For example 
\begin{equation}\label{eq:flow couplings}
\langle \langle \prod_{\alpha=1}^n  \lambda_\alpha(\mu) \rangle  \rangle = \int d\lambda_\alpha f(\{\lambda_\alpha\}) \prod_{\alpha=1}^n\lambda_\alpha(\mu)
\end{equation}
Given the beta function of couplings in this way, one can deduce the corresponding flow of ensemble by demanding that \eqref{eq:flow ensemble} be the same as \eqref{eq:flow couplings} which is equivalent to
\begin{equation}
\mu \partial_\mu f(\{\lambda_\alpha\},\mu) = - \partial_\alpha \bigl (\beta^\alpha(\lambda) f(\{\lambda_\alpha\}) \bigr)
\end{equation}
where $\mu$ is the RG scale. We can characterize these distributions by their moments, and follow their RG flows by the infinite set of $\beta$ functions
\begin{equation}\label{n-moment}
\beta_n(\{J^{a;\bar a}\})=\mu\partial_\mu\ \langle\langle \Pi_{i=1}^n J^{a_i;{\bar a}_i}\rangle\rangle_\mu  =\langle\langle \mu \partial_\mu ( \Pi_{i=1}^n J^{a_i;{\bar a}_i}) \rangle\rangle\,.
\end{equation}
We will refer to them as the distribution $\beta$ functions. Working with the distribution $\beta$ functions might be a bit counter-intuitive at times. For example, the simplest case is when $n=1$. Using the leading $\beta$ function, quadratic in $J^{a ; \bar a}$, in \eqref{BetaFunQuad} and performing the ensemble average with $\langle\langle J_{b;b'}\,J_{c;c'} \rangle\rangle \sim \frac{J^2}{N^3} \delta_{bc}\delta_{b'c'}$, the average $\beta$ function equals
\begin{equation}\label{neq1}
\beta_1= \beta( \langle\langle J^{a;{\bar a}}  \rangle\rangle_\mu ) =\mu \partial_\mu \langle\langle J^{a ; \bar a} \rangle\rangle_\mu \propto   \sum_{b,\bar b,c,\bar c} f_{abc}f_{\bar a \bar b \bar c} \langle \langle J^{b;\bar b}\,J^{c;\bar c} \rangle \rangle = 0.
\end{equation}
However, this does not mean the theory does not flow. 

In general, if we choose some functional form for the distribution, the RG flow can take us out of this subspace of functions. For example, working at finite $N$ in our model, the distribution does not remain Gaussian, or even symmetric under $J_{a;\bar a} \to -J_{a;\bar a}$, along the RG flow.  To see this, consider \eqref{n-moment} for $n=3$ 
\begin{equation}
\begin{aligned}
\beta_3=\mu\partial_\mu \langle\langle J^{a_1;{\bar a}_1} J^{a_2;{\bar a}_2} J^{a_3;{\bar a}_3}\rangle\rangle_\mu &\propto 
\langle\langle J^{a_1;{\bar a}_1} J^{a_2;{\bar a}_2}f_{a_3bc}f_{{\bar a}_3{\bar b}{\bar c}} J^{b;{\bar b}}  J^{c;{\bar c}} \rangle\rangle\ + \text{permutations} \\
& \propto f_{a_1a_2a_3}{\bar f}_{\bar a_1 \bar a_2 \bar a_3} \frac{J^4}{N^6}\,.
\end{aligned}
\end{equation}
Hence, $\beta_3\neq 0$ at finite $N$, which is not compatible with a gaussian distribution. We observe, though, that this departure from gaussianity is subleading in $N$, compared to the nominal scaling in which $J^{a,{\bar a}}\sim J/N^{3/2}$.

\paragraph{Flow of the couplings:}
The various $\beta_n$ are statistical averages of various couplings in the theory. In principle, all of them appear in different Callan-Symanzik equations and one needs to track the RG flow for all of them. At large $N$, and as long as the statistical symmetry is unbroken, some simplifications occur. Here, we will focus on some additional aspects of the flow generated by the leading term in the $\beta$ function given in \eqref{BetaFunQuad}. We will see this term gives vanishing contributions (in the large N limit) to some of the interesting $\beta_n$'s. 
\begin{itemize}
\item  First, the ensemble average of the $\beta$ function was already computed in \eqref{neq1} 
\begin{equation}
\beta_1=\langle\langle \beta_{aa'} \rangle\rangle =  \langle\langle \mu \partial_\mu( J^{a;\bar a} ) \rangle\rangle = 0.
\end{equation}
Note this result also follows from symmetry --  the $\beta$ function transforms in the $\text{ad}(\SO(N)_L) \times \text{ad}(\SO(N)_R)$ representation and $\SO(N)_L\times \SO(N)_R$ is restored after the ensemble average. Hence the averaged $\beta$ function must vanish. 

\item Although we find that the averaged $\beta$ function vanishes, it could be that there are large fluctuations around this mean value. To gain some intuition on what the physics may be for a given realisation, we compute the standard deviation of this averaged beta function. This is given by $\beta_{a,a'}^2$ (not summed over $a,a'$) 
\begin{equation}
\begin{split}
\langle \langle \beta_{a,a'}^2 \rangle \rangle = & \sum_{bb'cc'{\hat b}{\hat b}'{\hat c}{\hat c}'} f_{abc}f_{a'b'c'}f_{a{\hat b}{\hat c}}f_{a'{\hat b}'{\hat c}'}  \langle \langle J_{b;b'}J_{c;c'}J_{{\hat b};{\hat b}'}J_{{\hat c};{\hat c}'}\rangle \rangle \\ 
= & {J^4\over N^6} \sum_{bb'cc'{\hat b}{\hat b}'{\hat c}{\hat c}'} f_{abc}f_{a'b'c'}f_{a{\hat b}{\hat c}}f_{a'{\hat b}'{\hat c}'} 
\bigl( \delta_{b{\hat b}}  \delta_{b'{\hat b}'} \delta_{c{\hat c}}  \delta_{c'{\hat c}'} + \delta_{b{\hat c}}  \delta_{b'{\hat c}} \delta_{c{\hat b}}  \delta_{c'{\hat b}'} \bigr) \\
\sim & {J^4\over N^6} \sum_{bcb'c'} f_{abc}f_{abc}f_{a'b'c'}f_{a'b'c'}\sim {J^4\over N^4} 
\end{split}
\end{equation}
The associated nominal scaling is  
\begin{equation}
{\partial {J\over N^{3/2}} \over \partial \log \mu } \propto {J^2\over N^2},\ \implies {\partial J\over \partial \log \mu}\sim {J^2\over \sqrt{N}} 
\end{equation}
and hence to leading order in the large N expansion (with J held fixed), the leading term in the $\beta$ function should be taken to be zero.

\item If the ensemble remains gaussian to leading order in $N$, the only parameter characterising the ensemble is $J$ (and in any case, the standard deviation is the leading statistical moment). Hence it is interesting to study its flow by studying 
\begin{equation}
J^2(\mu) ={1\over N} \langle \langle \sum_{a{\bar a}} {J_{a;{\bar a}}}^2 \rangle\rangle_\mu\ .
\end{equation}
At this order of the $\beta$ function, i.e. quadratic order in the couplings, $\beta_2$ also vanishes
 \begin{equation}
\beta_2=\mu \partial_\mu \langle\langle J^{a_1 ; \bar a_1} J^{a_2 ; \bar a_2} \rangle\rangle_\mu \propto f_{a_1bc}f_{\bar a_1 \bar b \bar c}  \langle\langle J^{b ; \bar b} J^{c ; \bar c} J^{a_2 ; \bar a_2} \rangle\rangle + 1 \leftrightarrow 2 = 0\ ,
\end{equation} 
i.e., 
\begin{equation}
 \partial_{\log \mu} J^2 = \partial_{\log \mu } \frac{1}{N} \langle\langle J_{a{\bar a}}J^{a{\bar a}} \rangle\rangle 
 =0\ ,
\end{equation}

\end{itemize}
These are actually different measures why the leading quadratic term in the $\beta$ function \eqref{BetaFunQuad} does not contribute to the flow, due to cancellation between the different couplings. These arguments are correct in perturbation theory, or as long as the $\SO(N)_L\times \SO(N)_R$ is unbroken. They may be limited in teaching us about global issues of the RG flow because the latter is non-linear in the parameters. For example, if we fix the couplings at some high scale $\Lambda$, then since relevant couplings are turned on, then eventually an IR dynamical scale would be generated (at least at finite N). Computing this dynamical scale for a specific realisation, as a function of $J_{a;\bar a}$, and then averaging over the $J_{a;\bar a}$ is different from first averaging the $\beta$ function and then trying to deduce the flow's end.

\vspace{10pt}
\subsection{$\beta$ function from conformal perturbation theory }
\label{pform}

Conformal perturbation theory provides an expansion of the $\beta$ function in higher powers of the couplings $J_{a;\bar a}$.
We saw the quadratic terms are effectively zero at large $N$ in our models. Yet the distribution parameter $J^2$ still flows as we see explicitly in our computations \eqref{betagamma}. In this section, we argue this RG flow originates from the cubic terms in the $\beta$ function.

Before starting this discussion, let us comment on the status of these cubic terms, which are usually scheme dependent and hence non-universal. For example, if we redefine the couplings appearing in \eqref{BetaConf} by
\begin{equation}\label{schemedep}
{\tilde \lambda}_\alpha=\lambda_\alpha+ \sum_{\beta \gamma} A_{\alpha\beta\gamma}\lambda_\beta\lambda_\gamma + \dots
\end{equation}
the higher order terms in the $\beta$ function expansion will change. However,  we can argue that such terms are universal in schemes preserving the $\SO(N)_L \times \SO(N)_R$ statistical symmetry. To see this, notice that the redefinition \eqref{schemedep} would change the cubic term in the beta function as follows\footnote{Here $\dot \lambda \equiv \mu \partial_\mu \lambda$.}
\begin{equation}
\begin{split}
{\dot{\tilde\lambda}_\alpha} &= {\dot\lambda}_\alpha + 2A_{\alpha\beta\gamma}{\dot\lambda}_\beta\lambda_\gamma =C_{\alpha\beta\gamma}\lambda_\beta\lambda_\gamma+{\cal O}(\lambda^3)+2A_{\alpha\beta\gamma}C_{\beta\sigma\delta}{\lambda}_\sigma\lambda_\delta \lambda_\gamma\\
&= C_{\alpha\beta\gamma}{\tilde\lambda}_\beta{\tilde\lambda}_\gamma +{\cal O}(\lambda^3) - 2C_{\alpha\beta\gamma}A_{\beta\sigma\delta}{\lambda}_\sigma\lambda_\delta \lambda_\gamma + 2A_{\alpha\beta\gamma}C_{\beta\sigma\delta}{\lambda}_\sigma\lambda_\delta \lambda_\gamma\,,
\end{split}
\end{equation}
where $C_{\alpha \beta \gamma}$ are given in \eqref{CintermsofF}. If the redefinition preserves the $\SO(N)_L\times \SO(N)_R$, the matrix $A$ has to intertwine $\SO(N)_L \times \SO(N)_R$ to $\SO(N)_L \times \SO(N)_R \times \SO(N)_L \times \SO(N)_R$.  The only option is to take $A_{\alpha \beta \gamma} \propto C_{\alpha \beta \gamma}$ but then the cubic term in the $\beta$ function does not change. 

In the following, we work with such a set of schemes, dictated by symmetry and having a universal cubic term contribution to the $\beta$ function. We use the results in \cite{Gaberdiel:2008fn,Behr:2013vta}. The formula for the $\beta$ function evaluated to this order is given in equations (5.12) and (5.13) of \cite{Behr:2013vta} which when adapted to our notation gives \footnote{Our convention for the $\beta$ function has an extra sign compared to that of \cite{Behr:2013vta}. In what follows, we will raise and lower adjoint indices $a,b$ etc by $\delta^{ab}, \delta_{ab}$ respectively.}
\begin{equation}
\beta_{a{\bar a}}=-\pi\,f_{bca}\,{ f}_{\bar b\bar c\bar a}\,J^{b;{\bar b}} J^{c;{\bar c}} -\beta_{a{\bar a} b{\bar b}c{\bar c}d{\bar d}} J^{b;{\bar b}} J^{c;{\bar c}} J^{d;{\bar d}}
\label{betak}
\end{equation}
where 
\begin{equation}
\begin{aligned}
\beta_{a{\bar a}   b{\bar b}c{\bar c}d{\bar d}} &= \frac{\pi^2}{3!} \biggl(E_{abcd,{\bar a}{\bar b}{\bar c}{\bar d}}+{\bar E}_{abcd,{\bar a}{\bar b}{\bar c}{\bar d}}\biggr) \\
E_{abcd,{\bar a}{\bar b}{\bar c}{\bar d}}&=
(\delta_{ad}\delta_{bc}-\delta_{ac}\delta_{bd}){f}_{{\bar a}{\bar b}}^{\ \ \bar r}{f}_{{\bar r}{\bar c}{\bar d}}+ 
 (\delta_{ab}\delta_{cd}-\delta_{ad}\delta_{bc}){f}_{{\bar a}{\bar c}}^{\ \ \bar r}{f}_{{\bar r}{\bar d}{\bar b}}  \\
 & \quad + (\delta_{ac}\delta_{bd}-\delta_{ab}\delta_{cd}){f}_{{\bar a}{\bar d}}^{\ \ \bar r}{f}_{{\bar r}{\bar b}{\bar c}}\,, \\ 
{\bar E}_{abcd,{\bar a}{\bar b}{\bar c}{\bar d}}&= (\delta_{\bar a\bar d}\delta_{\bar b\bar c}-\delta_{\bar a\bar c}\delta_{\bar b\bar d}) {  f}_{{a}{b}}^{\ \ r}f_{{ r}{c}{d}} + (\delta_{\bar a\bar b}\delta_{\bar c\bar d}-\delta_{\bar a\bar d}\delta_{\bar b\bar c}) {  f}_{{a}{c}}^{\ \ r}f_{{ r}{d}{b}} \\
& \quad + (\delta_{\bar a\bar c}\delta_{\bar b\bar d}-\delta_{\bar a\bar b}\delta_{\bar c\bar d}) {  f}_{{a}{d}}^{\ \ r}f_{{ r}{b}{c}}\,. \\
&= E_{{\bar a}{\bar b}{\bar c}{\bar d}, a bc d}
\end{aligned}
\label{betak2}
\end{equation} 

We are interested in evaluating the ensemble average of the $\beta$ function of the mean squared couplings, i.e $\tilde \beta \equiv \langle \langle \beta (J^{a;\bar a}  J_{a;\bar a}) \rangle \rangle$. Assuming the ensemble to be gaussian at all scales gives
\begin{equation}\label{betafun}
\tilde \beta  = \sum_{a,\bar a}\mu\partial_\mu \langle \langle J_{a;{\bar a}} J_{a;{\bar a}} \rangle \rangle_\mu = \sum_{a,\bar a} \mu\partial_\mu {2J^2\over N^3} \delta^a_a \delta^{\bar a}_{\bar a} =  J  N (1 - N^{-1})^2 \mu\partial_\mu J\,.
\end{equation}
Whereas the usual definition of $\beta$ function of ensemble gives 
\begin{eqnarray}\label{RHS}
\tilde \beta &=& 2 \sum_{a,\bar a} \langle \langle J_{a; \bar a} \mu\partial_\mu J_{a;\bar a} \rangle \rangle = 2 \sum_{a,\bar a} \langle \langle J_{a; \bar a} \beta^{a \bar a}\rangle \rangle \\
&=& - 2 \beta_{a{\bar a} b_1{\bar b}_1 b_2{\bar b}_2 b_3{\bar b}_3} \langle \langle J^{a;{\bar a}} J^{b_1;{\bar b}_1} J^{b_2;{\bar b_2}} J^{b_3;{\bar b_3}}\rangle \rangle\\ 
\nonumber
&=& -{2\pi^2 \over 3!} \left( E_{ab_1b_2b_3,\bar a \bar b_1 \bar b_2 \bar b_3} + E_{\bar a \bar b_1 \bar b_2 \bar b_3,ab_1b_2b_3} \right) \times   {4 J^4 \over N^6}[  \delta^{ab_1}\delta^{b_2b_3}\delta^{{\bar a}{\bar b}_1}\delta^{{\bar b}_2{\bar b}_3}+ 1,2,3 \ \text{cyclic}] \\
&=& -{16 \pi^2 J^4 \over 3! N^6} \quad   E_{ab_1b_2b_3,\bar a \bar b_1 \bar b_2 \bar b_3}  \quad  [  \delta^{ab_1}\delta^{b_2b_3}\delta^{{\bar a}{\bar b}_1}\delta^{{\bar b}_2{\bar b}_3}+ 1,2,3 \ \text{cyclic}] \\
\nonumber
&=& 4 \pi^2 J^4 N (1 + {\cal O}(N^{-1}))\,,
\end{eqnarray}
where we used relevant identities satisfied by the structure constants described in appendix \ref{currents}. Notice that the above has the correct nominal scaling ${1 \over N^3}\times N^4$.  Comparing with \eqref{betafun}, we get
\begin{equation}
\beta(J) \equiv \mu \partial_\mu  J = 4 \pi^2 J^3
\end{equation}
This matches the $\beta$ function in \eqref{betagamma} obtained from direct computation using the Callan-Symanzik equation.

\paragraph{Flow of ensemble :} Finally, it is worth noting that this cubic term preserves the Gaussianity of the distribution of the couplings at large $N$.  To prove this we can show that the RG flow of any statistical average of $2n$ $J$'s is given by taking into account, at leading order in N, by the RG flow of each pair. i.e., Gaussianity implies that
\begin{equation}\label{Gaussian Claim}
\begin{split}
\mu \partial_\mu \langle\langle J^{a_1;{\bar a}_1}....J^{a_{2n};{\bar a}_{2n}} \rangle\rangle_\mu = \sum_{\sigma} \langle\langle \mu \partial_\mu ( J^{a_{\sigma(1)};{\bar a}_{\sigma(1)}}  J^{a_{\sigma(2)};{\bar a}_{\sigma(2)}}) \rangle\rangle 
\prod_{i=2}^{n}  \langle\langle   J^{a_{\sigma(2i-1)};{\bar a}_{\sigma(2i-1)}}  J^{a_{\sigma(2i)};{\bar a}_{\sigma(2i)}}) \rangle\rangle 
\end{split}
\end{equation}
where the permutation $\sigma \subset S_{2n}$ has $n$ 2-cycles - basically all possible Wick contractions. To see this, we evaluate the left hand side first
\begin{equation}
\mu \partial_\mu \langle\langle J^{a_1;{\bar a}_1}....J^{a_{2n};{\bar a}_{2n}} \rangle\rangle_\mu  = - \sum_{k=1}^{2n}\beta_{a_k \bar a_k b_1 \bar b_1 b_2 \bar b_2 b_3 \bar b_3} \langle \langle  J^{b_1;\bar b_1} J^{b_2;\bar b_2} J^{b_3;\bar b_3}   \prod_{i=1,i\ne k}^{2n} J^{a_i; \bar a_i}   \rangle\rangle\,.
\end{equation}
The evaluation of the right hand side has two kinds of terms. One in which two of the $J^{b_i;\bar b_i}$ contract among themselves. Notice these are exactly the terms captured by \eqref{Gaussian Claim}. Their nominal scaling is $N^{-3n}$. The second kind involves no contractions among the $J^{b_i;\bar b_i}$. Consider one such term below (where say $J^{b_i;\bar b_i}$ contracts with $J^{a_i;\bar a_i}$ for $i=1,2,3$ and $k>3$). We show it has subleading nominal scaling. Hence, it is negligible.  
\begin{equation}
\begin{aligned}
&  \beta_{a_k \bar a_k b_1 \bar b_1 b_2 \bar b_2 b_3 \bar b_3} \prod_{i=1}^3 \langle \langle  J^{b_i;\bar b_i}  J^{a_i;\bar a_i} \rangle\rangle 
 \langle \langle  \prod_{i=4,i\ne k}^{2n} J^{a_i; \bar a_i}   \rangle\rangle \\ 
 & \sim  N^{-{9  }} E_{a_k b_1b_2b_3,\bar a_k \bar b_1 \bar b_2\bar b_3} \prod_{i=1}^3 \delta^{a_i b_i} \delta^{\bar a_i \bar b_i} \langle \langle  \prod_{i=4,i\ne k}^{2n} J^{a_i; \bar a_i}   \rangle\rangle \\
  & \sim  N^{-9}\left[ \left( \delta_{a_k a_3} \delta_{a_1a_2} - \delta_{a_k a_2} \delta_{a_1 a_3} \right)  f_{\bar a_k \bar a_1}^{\ \ \ \ \bar r} f_{\bar r \bar a_2 \bar a_3 }  + 1,2,3 \ \text{cyclic} \right] \langle \langle  \prod_{i=4,i\ne k}^{2n} J^{a_i; \bar a_i}   \rangle\rangle 
\end{aligned}
\end{equation}
Since $\sum_{\bar r}f_{\bar a_k \bar a_1}^{\ \ \ \ \bar r} f_{\bar r \bar a_2 \bar a_3 }$ does not scale with $N$, we conclude the nominal scaling of this term is $N^{-3n -3}$, which is subleading compared to the $N^{-3n}$ captured by the gaussianity preserving term in \eqref{Gaussian Claim}.

\section{Discussion and Future Directions}
\label{disc}

We computed the 2-pt function for a set of $N$ 1+1 dimensional Majorana fermions with a random Thirring interaction. We obtained that the $\beta$ function is positive, which means that the theory is not renormalizable. The running is, however, weaker than the usual logarithmic running of marginal operators. 

As an effective theory below some scale, it is perfectly valid, and its RG exhibits some new features. This is the context in which the theory might be useful for black holes in AdS -- the model above might serve as a heuristic model for the degrees of freedom inside a black hole at some finite energy, and if we want to understand its surrounding weakly curved AdS outside the horizon, we need to couple it to additional "probe" degrees of freedom \cite{Berkooz:2016cvq}. The energy scale is therefore set by the temperature and the degrees of freedom of the black hole "never make it" to high energies.

There are some natural future directions
\begin{enumerate}
\item[1.] A better understanding of the solution to the Schwinger-Dyson equation: We treated the Schwinger-Dyson equations in an approximation in which the we solve the spatial "interaction" equation exactly, and the momentum "propagator" equation only approximately. This was convenient since the "interaction" equation is more non-linear. The fact that we were able to approximately solve the equations consistently for all values of $J$ for sufficiently low energies lends support to this approximation scheme. However, it would be worthwhile to see if the approximation can be justified, or tested on other examples.

\item[2.] Thermal partition function: One natural extension is to compute the thermal partition function.  We expect that we will be able to do it at low temperature, where we can use a long distance approximations of the type that we used above. 

\item[3.] 4-pt function: In the 0+1 dimensional SYK there was a truncation of the action, in terms of 2-pt functions, to a reparametrization invariant action, and there was no need to regulate the action in order to make sense of the 2-pt function. This enabled partial computation of the 4-pt function, although the final expression required a scale (or the re-introduction of the RG flow). Here we need to include the RG flow already at the level of the 2-pt function, which means the computation of the 4-pt function might be considerably more difficult. It is, however, the main indication whether the theory can be related to a bulk theory of any sort.

\item[4.] Renormalizability and stability: We would like to see whether the theory is renormalizable, and at what scales it develops a mass gap in the IR. The former might be easier in the bi-local description of the mode. To check for a mass gap one can look for poles in futher correlation functions. In the Gross-Neveu model, for example, one indication for the mass scale is a tachyonic pole in the 2-pt function for $\sigma$, which feeds into the a pole in the 4-pt function of the fermions (there is no pole in the 2-pt function of the fermion which is not renormalized when writing the theory with the $\sigma$ field).  To see this effect in our model, we again need to compute the 4-pt function of the fermions in the model.

\item[5.] The Thirring model is dual to Sine-Gordon models via bosonization dualities. This suggests that one can obtain an SYK like models with bosonic fields too, atleast in 2 dimensions. It is interesting to then explore if there are other interesting bosonic SYK models. 

\item[6.] Finally, it would be interesting to couple this model, at finite temperature, to additional "probe" degrees of freedom and examine whether it resembles a black hole in $AdS_3$.

\end{enumerate}

We hope to return to these issues in future work.

\vspace{20pt}
\subsection*{Acknowledgements}
We would like to thank Ofer Aharony and David Kutasov for illuminating discussions. The work of MB is supported by an ISF center of excellence grant (1989/14). PN gratefully acknowledges the support from International Centre for Theoretical Sciences (ICTS), India. The work of MR is supported by a Discovery grant from NSERC. The work of JS is supported by the Science and Technology Facilities Council (STFC) [grant number ST/L000458/1]. MB holds the Charles and David Wolfson Professorial chair of Theoretical Physics. JS's research was supported in part by Perimeter Institute for Theoretical Physics. Research at Perimeter Institute is supported by the Government of Canada through the Department of Innovation, Science and Economic Development and by the Province of Ontario through the Ministry of Research, Innovation and Science.

\vspace{20pt}
\begin{appendix}

\section{2d Euclidean CFT conventions}
\label{2dcft}

Given the Euclidean action $S[\nu,\bar \nu]$ for the Majorana fermions $\nu,\bar \nu$, the euclidean partition function is given by 
\begin{equation}
Z = \int [D\nu] [D \bar \nu] e^{-S[\nu,\bar \nu]}
\end{equation} 
Euclidean coordinates are labelled by $\vec x= (x_1,x_2)$ and momentum by $\vec p = (p_1,p_2)$. The definition of Fourier transform 
and its inverse are
\begin{equation}
{\cal F}(\vec p) = \int {d^2\vec x}\, e^{i \vec p . \vec x} f(\vec x)\,, \quad
f(\vec x) = \frac{1}{(2\pi)^2}\int {d^2\vec p}\,  e^{-i \vec p . \vec x} {\cal F}(\vec p)\,.
\end{equation}
Complex coordinates are defined as 
\begin{eqnarray}
z = x_1+i x_2\,, \quad \bar z = x_1 - i x_2 \quad \implies \quad \underbrace{dz d\bar z}_{d^2\!z} = 2 d^2\! \vec x
\end{eqnarray}
and similarly for momenta 
\begin{equation}
p = {p_1 - i p_2 \over 2}\,, \quad \bar p = {p_1 + i p_2 \over 2} \quad \implies \quad \underbrace{dp d\bar{p}}_{d^2\!p} = {d^2\vec p \over 2}
\end{equation}
One can check that  $z \bar z = |\vec x|^2$ and $|p|^2 \equiv 4 p \bar p = |\vec p|^2$
\begin{equation}
\vec p .\vec x = p z + \bar p \bar z\,, \quad \delta^2(\vec x) = 2 \delta(z) \delta(\bar z) \,, \quad \delta^2(\vec p) = {  \delta(p) \delta(\bar p)\over 2}
\end{equation}
The definition of Fourier transform in complex coordinates translates to 
\begin{equation}
\begin{split}
{\cal F}(p, {\bar p})[ f(z,\bar z)] &\equiv \int {d^2\!z \over 2} e^{i (p z + {\bar p} \bar z)} f(z,\bar z) \\
f(z, \bar z) &= {\cal F}^{-1}[{\cal F}(p, {\bar p})] \equiv \int {2d^2\!p \over (2\pi)^2} e^{-i (p z + {\bar p} \bar z)} {\cal F}(p, {\bar p})\,. \\
\end{split}
\end{equation}

\paragraph{Summary of Fourier transforms.} The main Fourier transforms used in the main text is given below.

\begin{eqnarray}\label{Fourier1}
	\int {d^2\!z \over 2} e^{i (p  z + \bar p  \bar z)} { \log^\alpha(\Lambda^2 z \bar z) \over z \bar z   } &=& { \pi  \over \alpha +1 }   \log^{\alpha+1} (\Lambda^2/|p|^2) \left( 1 +{\cal O}({ 1 \over \log(\Lambda^2 / |p|^2)  })\right)   \\
	\label{Fourier2}
	\int {d^2\!z \over 2} e^{i (p  z + \bar p  \bar z)} { \log^\alpha(\Lambda^2 z \bar z) \over z   } &=& { i \pi  \over \bar p }   \log^{\alpha} (\Lambda^2/|p|^2)\left( 1 +{\cal O}({ 1 \over \log(\Lambda^2 /|p|^2)  })\right)  \\
	\label{Fourier3}
	\int {d^2\!z \over 2} e^{i (p  z + \bar p  \bar z)} { \log^\alpha(\Lambda^2 z \bar z) \over z \bar z^2   } &=& { i \pi \bar p \over \alpha +1 }   \log^{\alpha+1} (\Lambda^2/|p|^2)\left( 1 +{\cal O}({ 1 \over \log(\Lambda^2 / |p|^2)  })\right)  
\end{eqnarray}

\vspace{20pt}
\section{A different regularisation of the main Fourier transform}
\label{alter}

In this appendix we provide an alternative regularisation for the integral \eqref{original}
\begin{equation}
  F_\alpha(|p|/\Lambda) = (2 \pi) \ 2^\alpha   \int_0^\infty {dr \over r} \log^\alpha(\Lambda r) J_0(|p| r)\,.
\end{equation}

Since the integral diverges at small $r$, we need to cut it off. It is natural to do so at $r=1/\Lambda$ (or larger values so that the logarithm does not become negative). This is another indication that the scale $\Lambda$ behaves like a UV cut-off. The resulting integral reduces to
\begin{equation}
F_\alpha(|p|/\Lambda) = \pi\, 2^{\alpha+1}\,\int_1^\infty \frac{dy}{y} \left(\log y\right)^\alpha J_0(|p|\,y/\Lambda)\,,
\label{Fintegral1}
\end{equation}
where we introduced the dimensionless variable $y=\Lambda\,r$.

Let us consider the $\alpha= n \in \mathbb{Z}$ to get some intuition. We are interested in studying the momentum dependence of this integral. For this purpose, define $\tau \equiv |p|/\Lambda$ and $t=\tau\,y$, so that \eqref{Fintegral1} equals
\begin{equation*}
I_\alpha(\tau) \equiv \frac{F_\alpha(|p|/\Lambda)}{\pi\, 2^{\alpha+1}}= \int^\infty_\tau \frac{dt}{t}(\log z/\tau)^\alpha\,J_0(t)\,.
\end{equation*}
Its derivative with respect to $\tau$ satisfies
\begin{equation}
\frac{dI_\alpha(\tau)}{d\tau} = -\frac{\alpha}{\tau} I_{\alpha-1}(\tau)\,.
\label{recurrence}
\end{equation}
Notice the contribution from the lower limit of integration does not contribute since the integrand vanishes at $t=\tau$ due to the logarithm.

To proceed, we use an integral appearing in \cite{abramowitz+stegun}
\begin{equation}
I_0(\tau) = \int^\infty_\tau \frac{J_0(t)}{t}\,dt = -\gamma - \log\frac{\tau}{2} - \sum_{k=1}^\infty (-1)^k \frac{(\tau/2)^{2k}}{2k\,(k!)^2}\,.
\end{equation}
This result is valid for any $\tau$. In the following, we will explore the physical regime corresponding to the IR, $\tau\equiv |p|/\Lambda \ll 1$, where the dominant contribution to $I_0(\tau)$ is captured by the logarithm and all the analytic terms are dropped.

It is easy to solve the recurrence relation \eqref{recurrence} when focusing on this dominant logarithmic contribution. Indeed, by induction, we can prove that the ansatz
\begin{equation}
I_\alpha(\tau) \sim \frac{(-1)^{\alpha+1}}{\alpha+1} \left(\log \frac{\tau}{2}\right)^{\alpha + 1}\,,
\end{equation}
solves \eqref{recurrence}. Notice that the integer nature of $\alpha$ was used in the induction step, so that $I_0(\tau)$ belongs to our series. 

The above derivation for $\alpha\in \mathbb{Z}$ also determines the logarithmic nature of the subleading contributions
\begin{equation}
F_\alpha(|p|/\Lambda)=\frac{\pi}{\alpha+1} \left(\log (\Lambda^2/|p|^2)\right)^{\alpha+1} \left(1+{\cal O}(1/\log(|p|/\Lambda)) \right)\,,
\end{equation}
in agreement with the result \eqref{GenFormFn} in the main text.

\vspace{20pt}
\section{Normalisation of currents}
\label{currents}

When matching the free fermionic notation with the $\SO(N)$ at level one, the currents are written in terms of adjoint indices $a$
\begin{equation}
  J^a(z) = \frac{1}{2}\sum_{i,j} \left(\nu^i\,t^a_{ij}\,\nu^j\right)(z)\,,
\end{equation}
with a similar expression for $\bar{J}^b(\bar{z})$ in terms of $\bar{\nu}^k$, where $a$ stands for a pair of integers $(kl)$ satisfying $1\leq k < l \leq N$ and the matrices $t^a_{ij}$ satisfy\footnote{Following the conventions in \cite{DiFrancesco:1997nk}.}
\begin{equation}
\begin{aligned}
 t^a_{ij} &\equiv i\left(\delta^k_i\delta^l_j - \delta^k_j\delta^l_i\right) \\
 \text{Tr}\left(t^at^b\right) &= 2\delta^{ab} \equiv 2\left(\delta^{ki}\delta^{lj}-\delta^{kj}\delta^{li}\right) \\
 \sum_a t^a_{ij}t^a_{kl} &= -\delta_{ik}\delta_{jl}+ \delta_{jk}\delta_{il}\,,
\end{aligned}
\end{equation}
where in the second line $a=(ij)$ and $b=(kl)$. Notice $\sum_a \delta^{aa} = \frac{N^2}{2}(1-N^{-1})$ due to the range of the indices $(kl)$. These matrices have a commutator $\left[t^a,\,t^b\right]= \sum_c if_{abc}\,t^c$ whose structure constants are explicitly given by
\begin{equation}
 f_{abc}  \equiv  f_{(ij)(kl)(mn)}=\delta_{mi}\left(\delta_{nl}\delta_{jk}-\delta_{nk}\delta_{jl}\right) + \delta_{mj}\left(\delta_{il}\delta_{nk}-\delta_{nl}\delta_{ik}\right) 
\end{equation}
It is easy to show these structure constants satisfy the identity
\begin{equation}
  \sum_{a,b} f_{abc}f_{abd} = 2(N-2)\delta_{cd}\,.
\label{soNid}
\end{equation}

Finally, the OPE between currents is
\begin{equation*}
  J^a(z)\,J^b(\omega) \sim \sum_c \frac{i\,f_{abc}\,J^c(\omega)}{(z-\omega)} + \frac{1}{2}\frac{\text{Tr}(t^at^b)}{(z-\omega)^2}\,,
\end{equation*}
with an analogous expression for the opposite chirality. These determine the 3-pt functions \eqref{CintermsofF}.

\end{appendix}

\vspace{10pt}
\bibliographystyle{JHEP}
\bibliography{bibl}

\end{document}